\documentclass[prb,preprint]{revtex4-1} 

\usepackage{xcolor}
\usepackage{amsmath}  
\usepackage{amsfonts} 
\usepackage{graphicx} 
\usepackage[utf8]{inputenc}
\usepackage{gensymb}
\begin{document}

\title{A student experiment on error analysis and uncertainties 
based on mobile--device sensors}

\author{Martín Monteiro}
\email{monteiro@ort.edu.uy} 
\affiliation{Universidad ORT Uruguay}

\author{Cecila Stari}
\author{Cecila Cabeza}
\author{Arturo C. Martí}
\email{marti@fisica.edu.uy}

\affiliation{Instituto de F\'{i}sica, Facultad de Ciencias,
  Universidad de la Rep\'{u}blica, Igu\'{a} 4225, Montevideo, 11200,
  Uruguay}

\date{\today}

\begin{abstract}
Science students must deal with the errors inherent to all physical
measurements and be conscious of the necessity to express their as a
best estimate and a range of uncertainty. Errors are routinely
classified as statistical or systematic. Although statistical errors
are usually dealt with in the first years of science studies, the
typical approaches are based on performing manually repetitive
observations. Here, based on data recorded with the sensors present in
many mobile devices a set of laboratory experiments to teach error and
uncertainties is proposed. The main aspects addressed are the physical
meaning of the mean value and standard deviation, and the
interpretation of histograms and distributions. Other activities focus
on the intensity of the fluctuations in different situations, such as
placing the device on a table or held in the hand in different ways
and the number of measurements in an interval centered on the mean
value as a function of the width expressed in terms of the standard
deviation. As applications to every day situations we discuss the
smoothness of a road or the different positions to take photographs
both of them quantified in terms of the fluctuations registered by the
accelerometer. This kind of experiments contributes to gaining a deep
insight into modern technologies and statistical errors and, finally,
to motivate and encourage engineering and science students.
\end{abstract}

\maketitle

\section{Introduction}
In many experimental situations when a measurement is repeated, for
example when we measure a time interval with a stopwatch, or the
distance at which a ball launched with a spring-loaded projectile
launcher falls or a voltage with a digital multimeter, successive
readings, under identical conditions, give slightly different
results. This occurs beyond the care we take to always launch the
balls in exactly the same way or to connect the components of the
circuit so that they are firmly attached. In effect, this phenomenon
is due to the fact that most measurements in the real world present
statistical uncertainties
\cite{taylor1997introduction,hughes2010measurements}.
When facing repeated observations with different results it is natural
to ask ourselves what is the most representative value and what is the
confidence that we can have in that value.  The International Standard
Organization (ISO) \cite{iso1995guide} defines the errors evaluated by
means of the statistical analysis of a series of observations as type
A in contrast with other, \textit{systematic}, sources of errors, type
B, whose evaluation is estimated using all available non-statistical
information such as instrument characteristics or observer's
individual judgment. In this work, we focus on the teaching of
statistical errors in the first years of engineering and science
studies using modern sensors.

The study of error analysis and uncertainties plays a prominent role
in the first years of all science courses. Perhaps the most
important message is to persuade students that any measurement is
useless unless a confidence interval is specified.  It is expected
that after finishing their studies, students are able to discuss
whether a result agrees with a given theory, or if it is reproducible,
or to distinguish a new phenomenon from other already known. With this
objective, various experiments are usually proposed in introductory
laboratory courses
\cite{4321968,mathieson1970student,fernando1976experiment,wibig2013hands,gan2013simple}.
These experiments usually involve a great amount of repetitive
measurements such as dropping small balls \cite{gan2013simple} or
measuring the length of hundred or thousands of nails using a vernier
caliper \cite{fernando1976experiment}.  The measurements obtained are
usually examined from the statistical viewpoint plotting histograms,
calculating mean values and standard deviations and, eventually,
compared with those expected from a known distribution, typically a
normal distribution.  Though these experiments are illustrative, most
of them are tedious and do not adequately reflect the present state of
the art.

The importance for their careers of a physicist being able to design a
measurement procedure, select the equipment or instruments, perform
the process and finally express the results as the best estimation and
its uncertainties has been remarked. However, recent studies
\cite{Sere1993learning,allie2003teaching,chimeno2005teaching}, suggest
that students lack these abilities. Several difficulties have been
described \cite{Sere1993learning}: the lack of understanding of the
need to make several measurements, or insight into the notion of
confidence interval or the ability to distinguish between random and
systematic errors.

Mobile devices such as smartphones or tablets which usually include
several sensors (accelerometer, magnetometer, ambient light sensor,
among others) appear as modern and versatile alternatives to deal with
statistical errors.  In fact, the use of smartphone sensors has been
proposed in many science experiments
\cite{vieyrafive,hochberg2018using}, ranging from experiments with
quadcopters \cite{monteiro2016using} to shadowgraph imaging
\cite{SETTLES20189}. The inevitable noise of the sensors, so annoying
in any measurement, can be used, however, favorably, to illustrate
basic concepts of statistical treatment of measurements. It is
possible, using these sensors, to acquire hundreds or thousands of
repeated values of physical magnitude in a few seconds that can be
analyzed in the mobile device or transferred to a PC. Thanks to their
sensitivity these values clearly display statistical fluctuations. In
this paper we propose a set of laboratory activities to teach error
analysis and uncertainties using modern technologies in a stimulating
approach. In the next Section we review some basic concepts about
error analysis, while in Section~\ref{sec:lab} we describe the
proposed activities. Finally, in Section~\ref{sec:con} we present the
summary and conclusion.

\section{Statistical errors}
In this work we focus on the teaching of statistical errors which due
to a multitude of causes are inherent to all physical measurements
\cite{taylor1997introduction}.  We assume that in a given experiment
an observation is repeated $N$ times under identical conditions
obtaining different results $x_i$, with $i=1,.. ,N$. It can be shown
that the best representative or \textit{estimate} of the set of values
is given by the mean value $\overline{x}$ defined as
\begin{equation}
 \overline{x}= \frac{1}{N} \sum_{i=1}^N x_i.
 \end{equation}
The deviation with respect to the mean value is identified with
$\epsilon_i= x_i - \overline{x}$. It can be shown that the mean value
defined as above minimizes the sum of the squared
deviations. Intuitively, it can be regarded as the
\textit{center-of-mass} of the set of the observations or equivalent
to the value \textit{closest} to all the other values.  In statistical
errors it is of interest to quantify the dispersion of the values
around the mean value or, informally, the\textit{ width of the cloud}.
The standard deviation defined as
\begin{equation}
  \sigma = \sqrt{\frac{1}{N-1}\sum_{i=1}^N (x_i - \overline{x})^2}
\end{equation}
can be seen as a measure of this dispersion. If the number of
observations, $N$, is large enough, $\sigma$ it is characteristic of
the set of all the possible observations and does not depend on the
specific set of observations.  In practice, the uncertainty in the
determination of a physical magnitude depends on the number of
repeated measurements we have done.

The standard error, or standard deviation of the mean, is defined as
$\sigma_x= \sigma/\sqrt{N}$ and it is demonstrated that it represents
the margin of uncertainty of the mean value obtained in a particular
set of measurements. The result of a specific measurement is usually
expressed in terms of the mean value and the standard error as
 \begin{equation}
  \overline{x} \pm \sigma_x  
 \end{equation}
representing the best estimate and the confidence in that value.  It
is worth highlighting that the standard deviation is related to the
degree to which an observation deviated from the mean value whereas
the standard error is an estimate of the uncertainty of the mean
value.  In a practical situation the standard error depends on the
number of measurements taken with $N^{-1/2}$. Then, given a set of $N$
measures the standard deviation gives an idea of the dispersion of an
ideal set of infinite measures while the standard error represents the
uncertainty of our set. This margin can be reduced by increasing the
number of measurements, however, the square-root implies that this
reduction is relatively slow.

It is an empirical fact that when the uncertainties of a continuous
magnitude do not have a preferred direction they follow a normal or
Gaussian distribution. The probability distribution function resembles
the well-known bell-shaped curve centered around the mean value
observed in many phenomena in natural and social sciences. The width
of the bell is given by the standard deviation, the inflection points
are located at $\overline{x} \pm \sigma$.

\section{A laboratory based on mobile devices}
\label{sec:lab}
The vast majority of smartphones and tablets have several built-in
sensors, in particular, triaxial accelerometers capable of measuring
the acceleration of the device in the three independent spatial
directions. Though it is possible to use all the components
simultaneously, here, for the sake of clarity, the following
experiments are based on the $z$ direction which is defined as
perpendicular to the screen. To access the values registered by the
sensors a specific piece of software or \textit{app} is necessary.

\begin{figure}[ht]
\begin{centering}
\includegraphics[width=0.99\columnwidth]{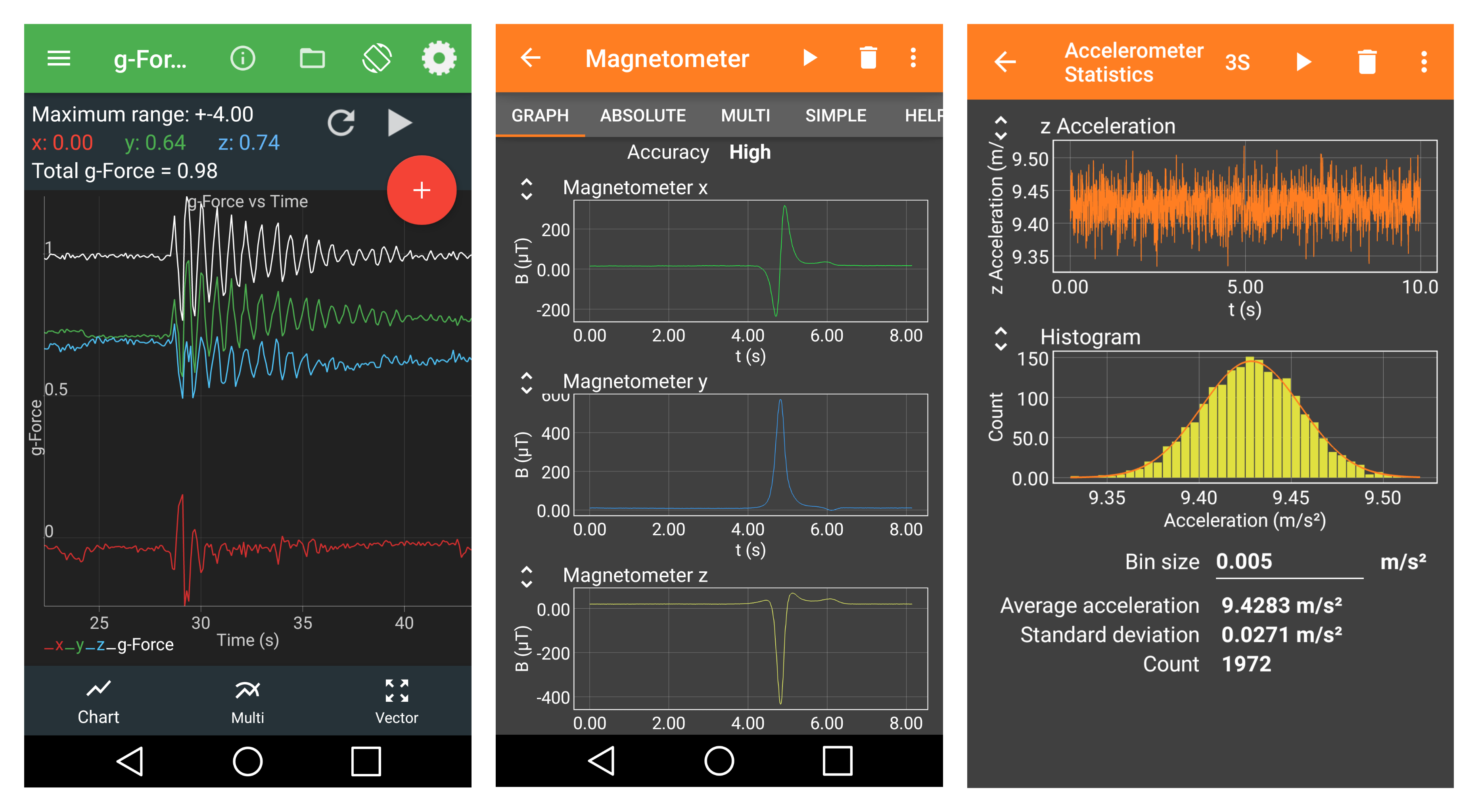}
\caption{\label{fig2apps} Screenshots of the most used \textit{apps}:
  Physics Toolbox suite (left) and Phyphox (center and right). The
  right panel shows a Phyphox screenshot of the experiment
  \textit{Statistical Basics} including a temporal series of the
  vertical component of the acceleration (top) and the corresponding
  histogram (bottom) overlapped with a Gaussian curve with the same
  mean and standard deviation indicated in the image.  }
\end{centering}
\end{figure}

From the many \textit{apps} available in the digital stores we
selected Physics Toolbox Suite \cite{physicstoolbox}, Androsensor and
PhyPhox \cite{Staacks_2018} whose screenshots are shown in
Fig.~\ref{fig2apps}.  Using these \textit{apps} it is possible to
select the relevant sensors, and to setup the parameters such as the
duration of the time series and the sampling frequency.  The
registered data can be analyzed directly on the smartphone screen or
transferred to the cloud and studied on a Personal Computer using a
standard graphics package.  Others useful characteristics present in
these \textit{apps} are the delayed execution and the remote access
via \textit{wi-fi} or browser. These capabilities allow the avoidance
of touching or pushing the mobile device when the experiments has
started.
 
\subsection{A first approach to fluctuations}
The first experiment consists of recording the fluctuations of the
vertical component of the accelerometer sensor with the mobile device
standing on a table, during a time lapse. In this experiment, and all
the described above, it is possible to use an app and download the
data or use the PhyPhox app or to choose in the menu the experiment
\textit{Statistical physics} which automatically displays temporal
series and histograms. In our case, we choose, unless stated
otherwise, a delay of $3$ s and register $a_z$ for $10$s.  The $3$ s
delay is important to avoid touching the device at the moment the
register starts and introducing spurious values.  The screenshot is
displayed in Fig.~\ref{fig2apps} (right). In this case the number of
measurements and the sampling period are $N=2501$ and $\Delta t=
0.004$ respectively.

\begin{figure}[ht]
\begin{centering}
\includegraphics[width=0.99\columnwidth]{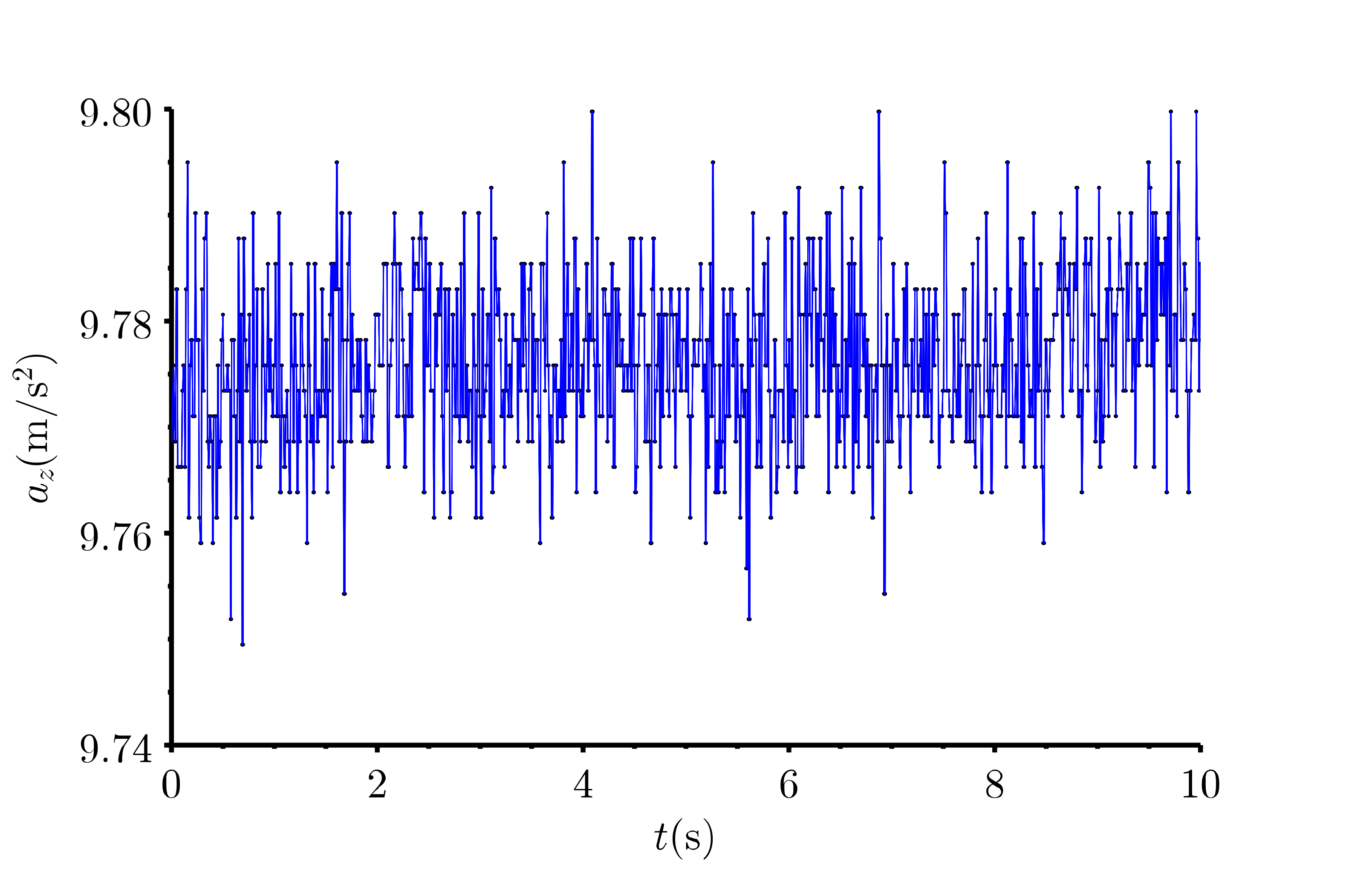}
\par\end{centering}
\caption{\label{figA} Temporal values of the $z$ component of the
  accelerometer while the smartphone is standing at rest on a table.
  The values registered by the sensor are indicated with small circles
  while the lines are guides for the eye.}
\end{figure}

Although the device is at rest on a horizontal surface, the $a_z$
values displayed in Fig.~\ref{figA} fluctuate steadily around a mean
value given by the gravitational acceleration $\overline{x} =
9,776$m/s$^2$ and a standard deviation $\sigma=0.008$m/s$^2$.  The
non-zero mean value is due to the fact that accelerometers are in fact
force sensors that cannot distinguish between the acceleration and the
gravitational field \cite{monteiro2014exploring,MONTEIRO2015}.  If,
instead of the acceleration sensor, the so-called \textit{linear}
acceleration pseudo-sensor were used, the measurements would fluctuate
around $0$ m/s$^2$.

\begin{figure}[ht]
\begin{centering}
\includegraphics[width=0.99\columnwidth]{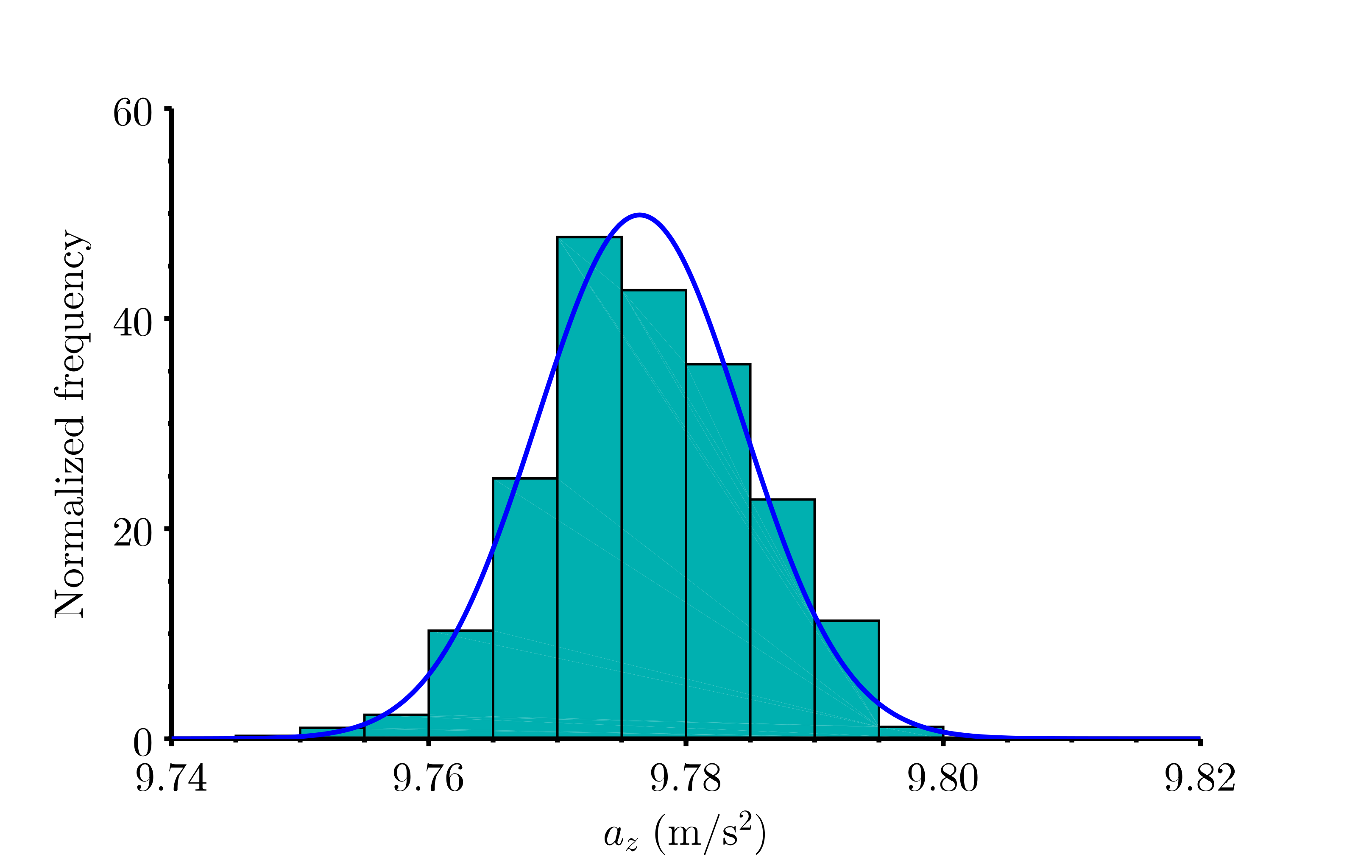}
\par\end{centering}
\caption{\label{fighisto} Histogram of the values from Fig.~\ref{figA}
  and a Gaussian curve with the same mean value, standard deviation
  and normalization}
\end{figure}

The corresponding histogram is displayed in Fig.~\ref{fighisto} with,
for the sake of comparison, a normal (Gaussian) curve with the same
mean value and standard deviation.  The vertical scale has been
adjusted so that the area under the normal curve and the sum of the
bins of the histogram are both equal to $1$.  From this figure, it can
be concluded that the histogram and the normal curve agree very
well. By increasing the number of samples $N$ and simultaneously
decreasing the width of the bins, it can be seen (not shown here) that
the agreement improves even more.

\subsection{Resolution in digital sensors}

\begin{figure}
\begin{centering}
\includegraphics[width=0.99\columnwidth]{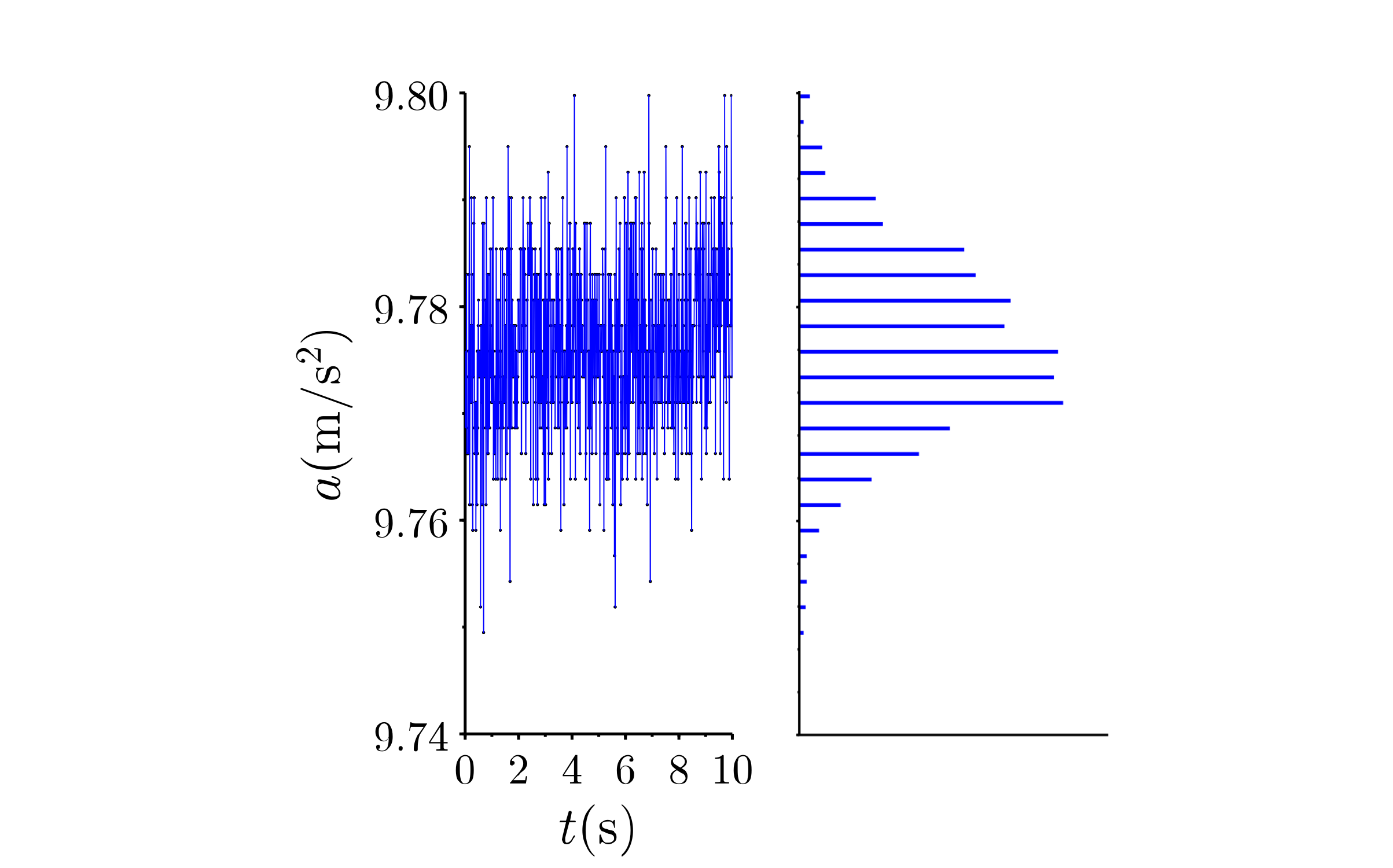}
\par\end{centering}
\caption{\label{figC} Discrete nature of the sensor data.  The left
  panel is similar to Fig.\ref{figA} but zoomed out in the horizontal
  axis to emphasize the discrete nature of the accelerometer
  values. The right panel shows the same values in a \textit{layed
    down} histogram with the same vertical scale.}
\end{figure}

It can be noticed in Fig.~\ref{figA} that the sensor values display a
clear regularity, the ordinates do not take arbitrary continuous
values but only a discrete set. This is more evident in
Fig.~\ref{figC} where, in the left panel, the horizontal axis of
Fig.~\ref{figA} has been zoomed out and, in the right panel, a
\textit{layed down} histogram with the same values is shown.  The
difference between the discrete values in the vertical axis is the
resolution of the instrument, that is, the minimum difference that the
sensor can register. This is typical of digital instruments, where a
continuous magnitude (such as acceleration, in this case) is
transformed by a sensor into an analog electrical signal, which is
transformed by an analog-to-digital converter (ADC) into a digital
signal which can only take certain discrete values. The acceleration
sensor of the Samsung S7 is a K6DS3TR, as shown in
Table~\ref{tabsensor}. The resolution given by the manufacturer
(sometines it appears incorrectly as accuracy), is $\delta =
0.0023942017$ m/s$^2$, which, as can be seen in Fig.~\ref{figC},
corresponds exactly to the difference between the groups of
acceleration values.

\begin {table}[h]
\begin{center}
\begin{tabular}{|c|c|c|c|}  
\hline 
Phone &	Sensor	&  Range (m/s$^2$)	&  Resolution (m/s$^2$)    \\  \hline 
Samsung Galaxy S7	 &K6DS3TR&	78,4532	&0,0023942017   \\ 
LG G3	&LGE&	39,226593	&0,0011901855   \\ 
Nexus 5	&MPU-6515	&19,613297&	0,0005950928   \\ 
 iPhone 6&	MPU-6700	& &	   \\ 
Samsung J6+&	LSM6DSL&	39,2266 &	0,0011971008   \\ 
Xiaomi Redmi Note7	&ICM20607	&78,4532&	0,0011901855   \\ 
Samsung Galaxy S9	&LSM6DSL	&78,4532&	0,0023942017   \\ 
\hline
\end{tabular}  
\caption{\label{tabsensor} Sensor characteristics of the devices used
  in the different activities obtained with the Androsensor app. In
  the case of the iPhone the manufacturer does not provides this
  information.}
\end{center}
\end {table}

The resolution of the sensor can be related to other important
characteristic of the digital sensors. One is the range of the sensor,
$R$, corresponding to the difference between the maximum and minimum
value that it is capable of measuring.  The maximum number of
different values that the sensor can register is $2^n$ where the $n$
is known as the number of bits of the DAC. Resolution is simply the
quotient between the range and the total number of different values,
that is,
\begin{equation}
 \delta = \frac{R}{2^n}.
\end{equation}

In the sensor used in this experiment Table \ref{tabsensor} shows that
the accelerometer used in this case can measure a maximum acceleration
of $78.4532$ m/s$^2$.  Since it registers not only positive measures,
but also negative accelerations, the range turns out to be twice the
maximum value, that is, $R = 156.9064$ m/s$^2$. Therefore it can be
determined that this sensor is capable of measuring $R/ \delta =
65536$ different values and since $65536 = 2^{16}$, this means that it
is a $16$-bit sensor. These characteristics can be easily verified on
the datasheets of the sensors.

\subsection{Different noise intensities}

\begin{figure}
\begin{centering}
\includegraphics[width=0.99\columnwidth]{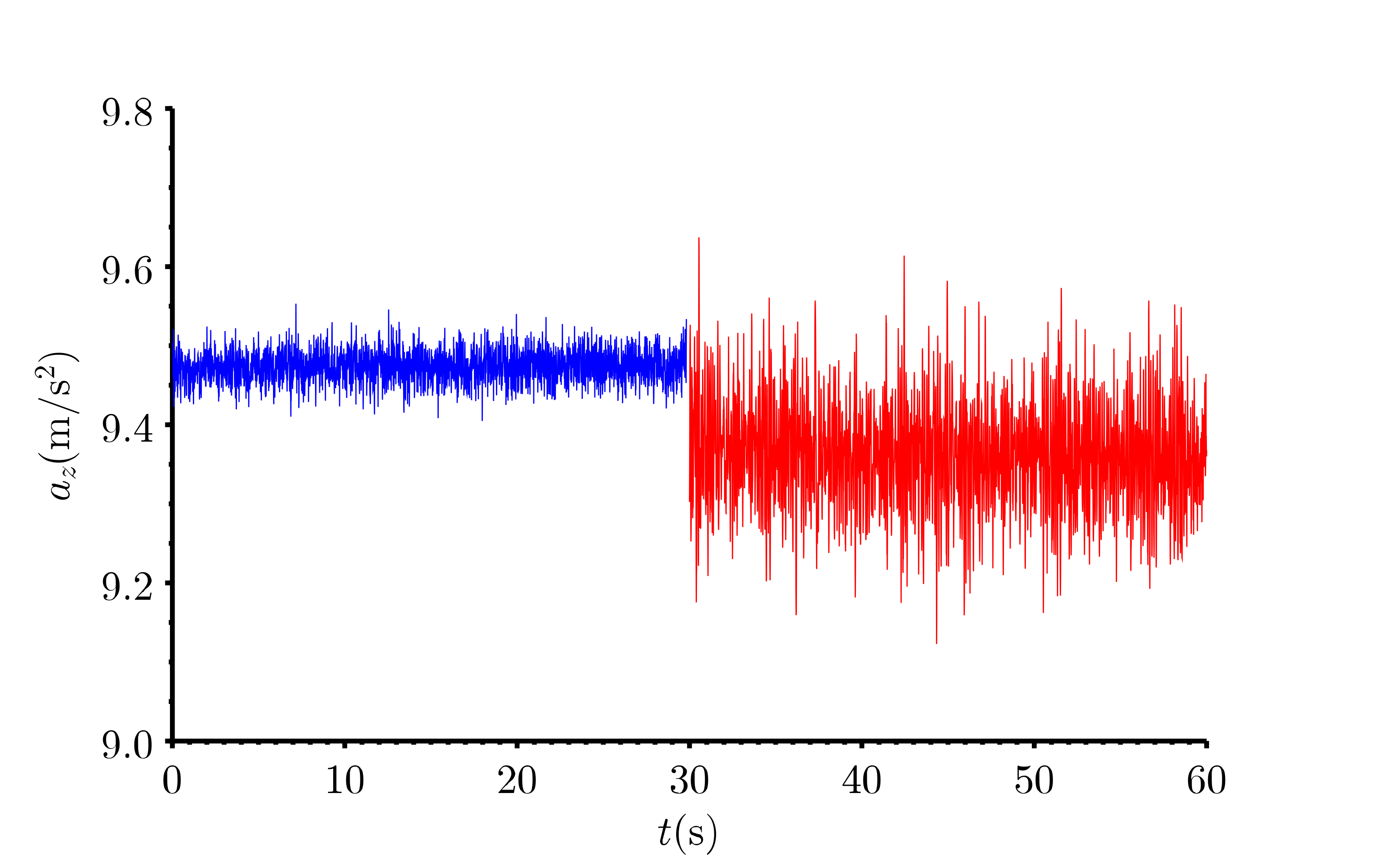}
\par\end{centering}
\caption{\label{figseriedoble} Temporal series of the acceleration
  values on the two different situations: first, the mobile device is
  on a table (blue lines) and secondly, held in a hand (red lines). }
\end{figure}

\begin{figure}
\begin{centering}
\includegraphics[width=0.99\columnwidth]{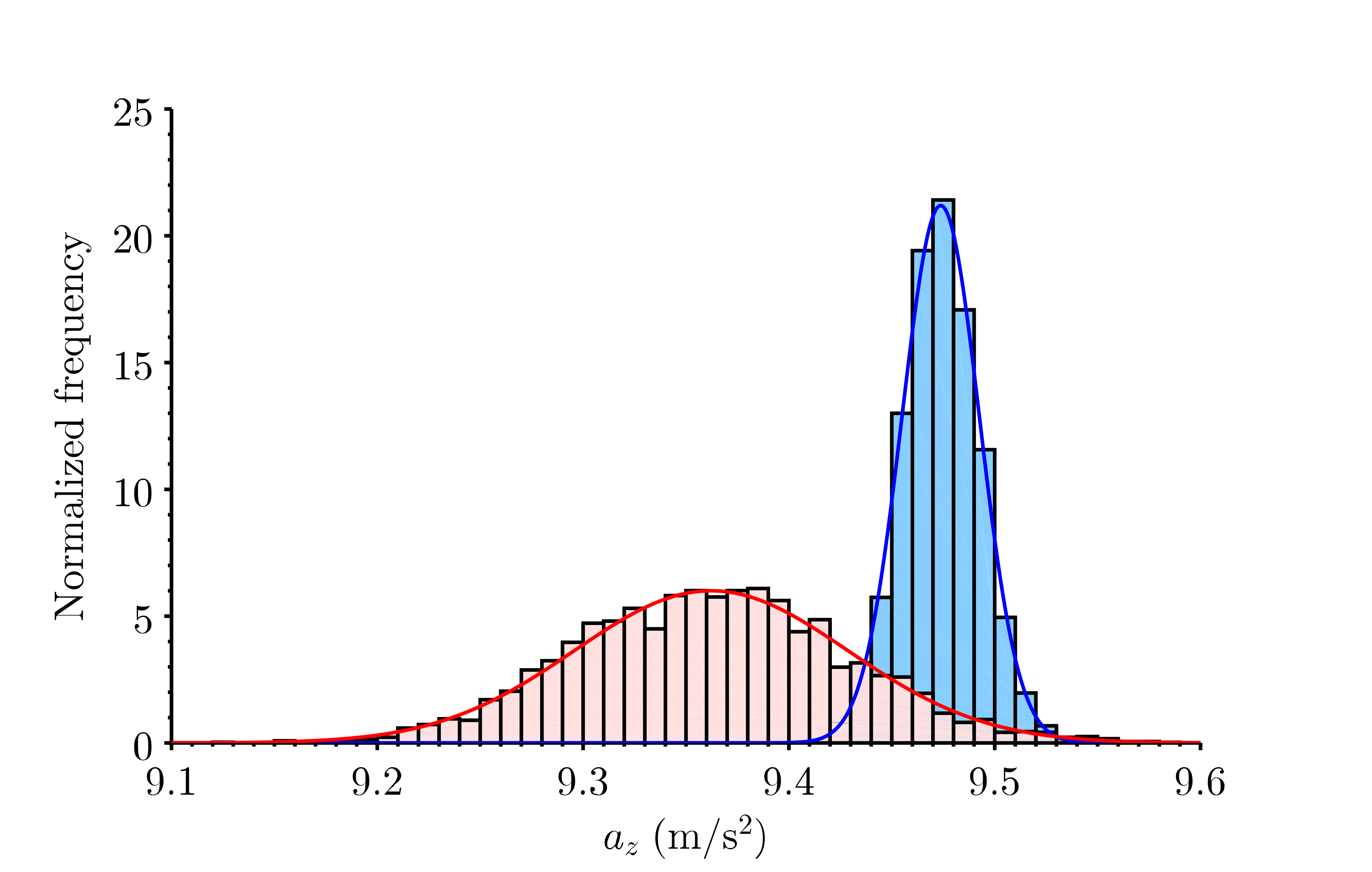}
\par\end{centering}
\caption{\label{figcomparative}Comparison between different noise
  intensities with the mobile device steady on a table (blue) or held
  in a hand (red). Data is the same displayed in the temporal series
  of Fig.~\ref{figseriedoble}. Continuous lines are Gaussian curves
  with the same mean values ($\overline{x}_{blue}=9.474$ m/s$^2$ and
  $\overline{x}_{red}=9.362$ m/s$^2$), standard deviations
  ($\sigma_{blue}=0.019$ m/s$^2$ and $\sigma_{red}=0.066$ m/s$^2$) and
  normalization factor corresponding respectively to the histograms of
  the same color.}
\end{figure}

In order to gain insight into the role of noise in different
situations in this experiment two sets of data are considered.  In the
first the smartphone is steady on a table and in the other the device
is held in the hand of the experimenter.  In Fig.~\ref{figseriedoble}
both temporal series are shown while in Fig.~\ref{figcomparative} the
corresponding histograms are displayed.  Moreover, histograms are
overlapped with normal curves with their respective mean values and
standard deviations.

It is clearly appreciated that the dispersion of data, quantified by
the standard deviation, is larger when the smartphone is held in the
hands than when the device is on the table. It is also noticeable in
both cases that normal curves agree very well with the histograms.
This activity can be translated to other settings. In particular, this
is one the basic mechanisms of seismographs.

\subsection{Number of observations in a given interval}

In general, the fundamental property of distributions is that the area
under one sector of the curve represents the probability that a new
measurement falls within this interval. In the case of normal curves,
it is usual to take intervals centered around the mean value and the
width in terms of the standard deviation. Then, it is shown that 68\%
of the observations will be in the "$\sigma$" interval, this is the
interval between $\overline{x} -\sigma$ and $\overline{x} + \sigma$,
\begin{equation}
P(\overline{x} - \sigma < x < \overline{x}+ \sigma) =
\int_{\overline{x} - \sigma}^{\overline{x} +\sigma}f(x) dx = 0.682...
\end{equation}
Similarly, the intervals "$2 \sigma$," "$3 \sigma$," and "$4 \sigma$"
concentrate, respectively, 95.4\%, 99.7\%, and 99.9\% of the
observations. This is a characteristic of normal distributions,
\textit{i.e.}, almost all the observations are concentrated around a
few "sigmas" around the mean value and graphically the curve is
relatively stretched.
 
\begin{figure}[ht]
\begin{centering}
\includegraphics[width=0.99\columnwidth]{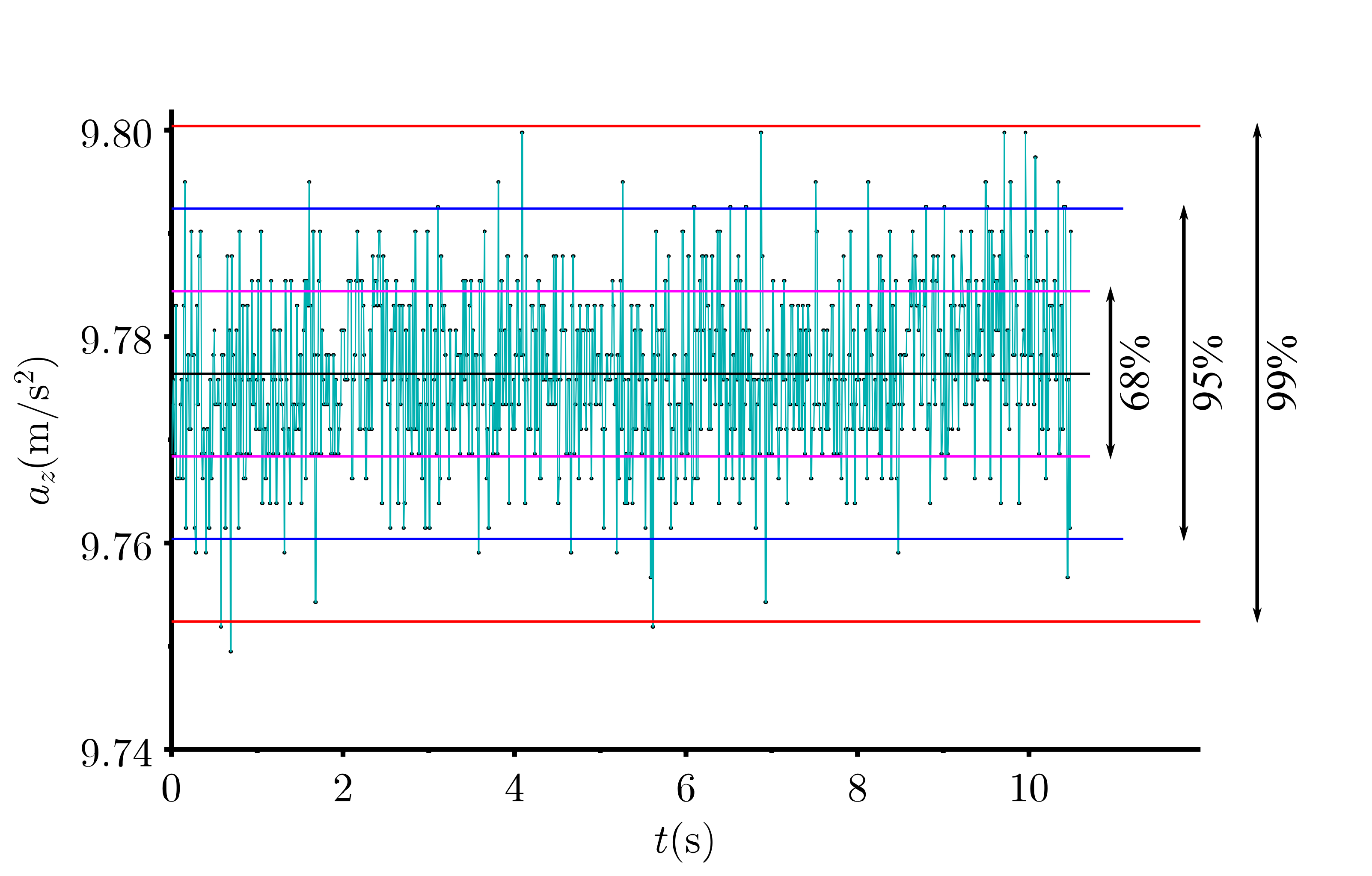}
\end{centering}
\caption{\label{figD4} Temporal series indicating the vertical
  intervals in term of units of the standard deviation $\sigma$.}
\end{figure}

To illustrate this phenomenon, Fig.~\ref{figD4} shows the temporal
series of Fig.~\ref{figA} with horizontal lines indicating the
$\sigma$ intervals.  It is evident from this figure that most of the
values concentrate around the mean value and a few $\sigma$
intervals. To quantify this relation, two experiments with different
noise intensities (on the floor and on an aircraft) are described in
Table~\ref{tab:sigma}. In this table the number of observations in a
given interval are compared with the expected values according to the
normal distribution.  It can be seen that the expected percentages are
similar to those according to a normal distribution.

An interesting point is to express these ranges in terms of the
resolution of the sensor. In this way it is noted that 68\% of the
measurements are within a radius interval equal to 3 times the
resolution. On the other hand, 99\% of the measurements are within a
radius interval equal to 10 times the resolution of the sensor.
\begin {table}[ht]
\begin{center}
\begin{tabular}{|c|c|c|c|c|}  
\hline \multicolumn{5}{|c |}{Experiment 1, $N=2098$, $a_z=(9.776\pm
  0.008)$ m/s$^2$ } \\ \hline Interval & Theo. (\%)& Theo. & Exp.(\%)
&Exp.  \\ $\overline{x}\pm \sigma$ &68.2 & 1431 & 70.0& 1468
\\ $\overline{x}\pm 2 \sigma$& 95.4 & 2001 & 95.5 &2003
\\ $\overline{x}\pm 3 \sigma$ &99.7 & 2092 & 99.6 &2090 \\ \hline
\multicolumn{5}{|c |}{Experiment 2, $N=1501$, $a_z=(9.65\pm 0.29)$
  m/s$^2$ } \\ \hline Interval & Theo. (\%)& Theo. & Exp.(\%) &Exp.
\\ $\overline{x}\pm \sigma$ &68.2 & 1024& 66.8 & 1003
\\ $\overline{x}\pm 2 \sigma$& 95.4 & 1432 & 94.7 &1422
\\ $\overline{x}\pm 3 \sigma$ &99.7 & 1497 & 99.7 &1497 \\ \hline
\end{tabular}  
\caption{\label{tab:sigma} Measurements in a given interval. Expected
  number of values according to a normal distribution and to the
  experimental results, respectively blue and red, displayed in
  Figs.~\ref{figseriedoble} and \ref{figcomparative}.}
\end{center}
\end {table}

\subsection{Optimal number of measurements}

Accuracy and precision are different concepts
\cite{hughes2010measurements}.  On the one hand, the precision of a
measurement, related to the random errors, is characterized by the
dispersion of the values, i.e. the standard deviation.  The smaller
$\sigma$, the less dispersion and therefore, the greater the
precision.  On the other hand, accuracy is related to systematic error
and quantified according to the agreement with an expected value.  As
mentioned above, in observations under identical and independent
conditions, the standard deviation does not change considerably with
the number of observations $N$. In contrast, the standard error,
giving account of the range of confidence in the estimation of the
mean value in a particular set of measurements decreases with
$N^{-1/2}$.  In Table.~\ref{tabsigmaN} the standard deviation and the
standard error are shown for different set of observations with
different $N$.  It is clear from that data, as mentioned above,
$\sigma$ is nearly constant while $\sigma_x$ clearly decreases.

\begin{table}
 \begin{center}
\begin{tabular}{|c|c|c|c|} \hline 
$N$ & $g$ & $\sigma$ & $\sigma_x$\\ \hline 
563 & 9,493 & 0,020 & 0,00085\\ 
1156 & 9,487 & 0,019 & 0,00054\\
1746 & 9,478 & 0,018 & 0,00044\\
2348 & 9,469 & 0,019 & 0,00039\\
2941 & 9,466 & 0,020 & 0,00036\\
3535 & 9,464 & 0,019 & 0,00032\\
4166 & 9,462 & 0,019 & 0,00029\\
4733 & 9,464 & 0,019 & 0,00027\\
5327 & 9,465 & 0,019 & 0,00026\\
5919 & 9,464 & 0,020 & 0,00026 \\ \hline
\end{tabular}
\caption{\label{tabsigmaN} Mean value, standard deviation and standard
  error as a function of the number of measures.}
\end{center}
\end{table}

As the decrease of the standard error with $N$ is slow, an important
question in practical situations is about the optimal number of
observations $N_{\mathrm{opt}}$.  Indeed, if we could repeat the
measurements infinite times we could achieve a perfect knowledge of
the best estimate and, accordingly, the standard error would
vanish. In fact, in addition to the statistical errors, type B errors
must be taken into account. In absence of other sources of systematic
errors, the optimal number of observations is defined when the
standard error is equal to the resolution of the digital instrument
$\sigma_x = \sigma /\sqrt{N} \sim \delta $. In the experiment,
depicted in Table~\ref{tabsigmaN}, given the resolution of S7 model
$0.0012$ m/s$^2$, the optimal number is $N_{\mathrm{opt}} \sim 250.$

\subsection{The best position to take a photograph with a cell phone}
An interesting experiment is to study the intensities of the
fluctuations depending on the way in which the experimenter holds
his/her device.  This activity can be adapted to be proposed as a
challenge to a group of students consisting in trying to hold the
device as steadily as possible.  Another possibility, not recommended
by the authors, is, similarly to Ref.~\cite{zaki2020study}, to study
the fluctuations of the gait of a pedestrian as a function of the
alcohol beverage intake.

The steadiness of the device is quantified by the standard deviation
of a given temporal series.  In Table~\ref{tab:position} we display
the intensities of the fluctuations in different positions. It is
evident from these values that keeping the device close to trunk
represents a more stable position.
 
\begin {table}[ht]
\begin{center}
\begin{tabular}{|l|l|l|l|l|}
\hline 
Smartphone position & N & $\sigma_\mathrm{G3}$ (m/s$^2$) & N& $\sigma_\mathrm{XR7}$ (m/s$^2$)\\ \hline
On the table      &1746 & 0.0184 & 2407 & 0.0052\\
Close to the body &1190 & 0.067  & 2502 & 0.1030\\
Selfie position   &1190 & 0.1206 & 2512 & 0.1413 \\ \hline
\end{tabular}  
\caption{\label{tab:position} Standard deviation of $a_z$ of two
  smartphone models in different positions.}
\end{center}
\end {table}

\subsection{The smartphone as a way to assess  road quality}
Recently, smartphones' sensors were proposed to assess road quality
\cite{harikrishnan2017vehicle}. In this activity, which can be
performed outdoors, students can assess the quality of a road. A means
of transport, in this case a car, under similar conditions (speed) is
employed, but other possibilities, such as a bike, are equally
feasible.  In Table~\ref{tab:roads} the intensities of the
fluctuations traveling by car on different roads are listed. To get an
insight of the fluctuations due to the road in the first row the noise
with the car stopped and engine idle is indicated. Just for the sake
of comparison a similar measurement but in a flying aircraft is
included.

\begin {table}[ht]
\begin{center}
\begin{tabular}{|l|l|l|l|l|} \hline 
 Situation & N & $\sigma_\mathrm{G3}$ (m/s$^2$) & N& $\sigma_\mathrm{XR7}$ (m/s$^2$)\\ \hline
Engine idle     &1181 & 0.3818 & 4984 & 0.0352\\
Smooth pavement &1200 & 1.3487 & 4974 & 0.5642\\
Stone pavement  &   -  &  -    & 4952 & 1.1491\\ 
Aircraft        & 1999 & 0.4374& -    & - \\ 
\hline
\end{tabular}  
\caption{\label{tab:roads} Assessment of the quality of different
  roads. Standard deviation of $a_z$ while the device is on the floor
  of the car with the screen orientated upwards.}
\end{center}
\end {table}

The intensity of the fluctuations depends on the specific sensor but exhibits
in all cases the same trends mentioned above. To summarize the results,
all the intensities of the fluctuations using the different built-in sensors 
in several situations are depicted in Fig.~\ref{figcomparative}.

\begin{figure}
\includegraphics[width=0.99\columnwidth]{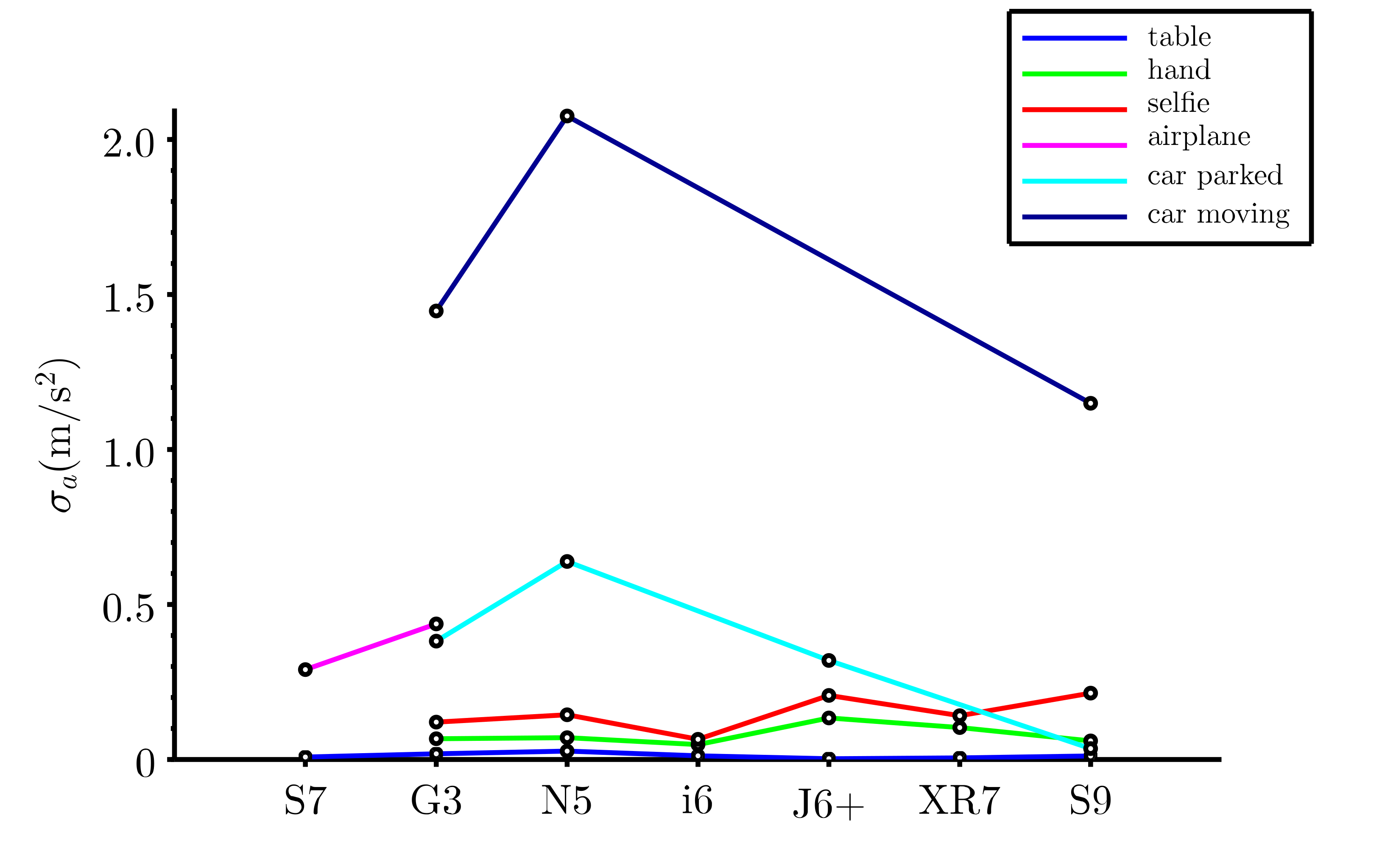}
\caption{\label{figsigmapos} Comparative table of the standard
  deviation $\sigma$ for different mobile devices in different
  activities as a function of the different models (see
  Table~\ref{tabsensor}).}
\end{figure}

\section{Conclusion}
\label{sec:con}

The main conclusion is that modern mobile-device sensors are useful
tools for teaching error analysis and uncertainties. In this work we
proposed several activities that can be performed to teach
uncertainties and error analysis using digital instruments and the
builtin sensors included in modern mobile devices. It is shown that
the distribution of fluctuations obeys normal (Gaussian)
statistics. Its main characteristics --mean, standard deviation,
histograms-- are analyzed. The role of noise intensity, spreading or
narrowing the normal bell-shapped curve is revealed. The width of the
distribution in terms of units of the standard deviation can be
related to the number of measurements in a given interval.  Holding
the mobile in different ways also gives an idea of how firmly it is
held.  In this approach, the lengthy and laborious manipulations
necessary in traditional approaches based on repetitive measurements,
are avoided allowing teaching to focus on the fundamental
concepts. These experiments could contribute to motivating students
and to showing them the necessity of considering uncertainty analysis.

\section*{Acknowledgment}
The authors would like to thank grant Fisica Nolineal (ID 722)
Programa Grupos I+D CSIC 2018 (UdelaR, Uruguay).

\bibliography{/home/arturo/Dropbox/bibtex/mybib}

\bibliographystyle{unsrt}

\end{document}